\documentstyle[12pt]{article}

\newcommand{\be}{\begin{equation}}
\newcommand{\ee}{\end{equation}}
\newcommand{\ov}{\overline}

\begin{document}
$~~$
\vspace{15mm}

\centerline{\Large \bf Universal scheme of minimal reduction}
\centerline{\Large \bf of usual and dual N=1,D=10 supergravity}
\centerline{\Large \bf to the Minkowsky space}

\bigskip

\centerline{{\bf K.N.Zyablyuk}\footnote{The research described in this
publication was made possible in part by the grant no. MOY000 from
the International Science Foundation.}}
\bigskip
\centerline{\it Institute of Theoretical and Experimental Physics,}

\bigskip

\begin{abstract}
The reduction from N=1, D=10 to N=4, D=4 supergravity with the
Yang-Mills matter is considered. For this purpose a set of constraints
is imposed in order to exclude six additional abelian matter
multiplets and conserve the supersymmetry.
We consider both the cases of usual and dual
N=1, D=10 supergravity using the superspace approach. Also the effective
potential of the deriving theory is written.
\end{abstract}

\section{Introduction}

The action of the N=1, D=10 supergravity is supposed to be the
effective  low-energy limit of the superstring action where
all the massive degrees of freedom are integrated out \cite{GSW}.
All the terms of this action may be characterized by the number
$$     n=N_\partial +\,{1\over 2}\,N_f                   $$
where $N_\partial$ and $N_f$ are numbers of derivatives and
fermions. The minimal supergravity has only the terms with $n=2$.
Usual \cite{CM} and dual \cite{GN} supergravity are equivalent
each other at the minimal level. But requirement of anomalies
cancellation implies that the Chern-Symons term must be presented
in the field-strength $H$ of the usual supergravity \cite{GS}. It leads to
appearance of terms with higher $n$ in the lagrangian. Here
the dual supergravity becomes more preferable because there is
a good reason to believe \cite{T2} that only the terms with $n=4$
must be added to the dual supergravity lagrangian while the usual
supergravity lagrangian turns out to be the infinite series
in $n$. We are not able to take into account terms with $n>2$ now
because they have not yet written explicitely but this work is
on the way.

We suppose that the ten-dimensional space-time $M^{10}$ is the
product ${M^4 \times K}$ of the Minkowsky space $M^4$ and some
internal manifold $K$ with unknown structure. It means that
every vector $V^M$ in $M^{10}$ decomposes into vector $V^\mu$
+ 6 scalars $V^m$ in $M^4$ (and vice versa in $K$) and every
Majorana-Weyl spinor in $M^{10}$ decomposes into 4 Majorana spinors
in $M^4$. So one can write down how the N=1,D=10 supergravity
degrees of freedom disintegrate under the reduction
$D=10\rightarrow D=4$ (in brackets the fields carried them are
denoted):
$$
\begin{array}{lrcl}
\;\;N=1 & D=10 \;\;& \rightarrow & \;\;D=4 \\
\;\;\mbox{Fermions:}& & & \\
\mbox{gravitino} &
56({\psi_M}) & = & \underline{4\times 2(\psi_\mu)}+24\times 2(\psi_m)\\
\mbox{dilatino} &
8(\chi) & = & \underline{4\times 2(\chi)}\\
\;\;\mbox{Bosons:}& & & \\
\mbox{graviton} &
35({E_M}^A) & = & \underline{2({e_\mu}^\alpha)}+
\underline{6\times 2({\cal B}_\mu^m)}+21({e_m}^a)\\
\mbox{dilaton} &
1(\phi) & = & \underline{1(\phi)}\\
\mbox{2-potential} &
28(B_{MN}) & = &\underline{1(B_{\mu\nu})}+
6\times 2(B_{m\mu})+15(B_{mn}) \\
\;\;\mbox{(usual)}& & & \\
\mbox{6-potential} &
28(M_{M_1 \ldots M_6}) & = & \underline{1(M_{m_1 \ldots m_6})}+
6\times 2(M_{\mu m_1 \ldots m_5})+\\
\;\;\mbox{(dual)}& & &\qquad\qquad\qquad +15(M_{\mu\nu m_1 \ldots m_4}) \\
\end{array}
$$
(for notations see below). It is easy to see that the underlined
fields constitute the multiplet of the N=4, D=4 supergravity.
All other fields are put together into 6 abelian matter multiplets.
Our aim here is to eliminate the abelian multiplets in order to
obtain the N=4, D=4 supergravity as a part of the N=1, D=10
supergravity where 6 coordinates $y^m$ are compactified on $K$.
We don't know whether these multiplets are essential in the
low-energy limit; but, definitely, the features of the theory
become more simple to analysis without them and they always can
be taken into account as a perturbation to the N=4, D=4 supergravity.

In usual N=1, D=10 supergravity without additional Yang-Mills (YM)
matter the separation of the abelian multiplets from pure N=4,D=4
supergravity has been realized in \cite{C1} and some attempts
to analise the dual case were made in \cite{C2}. In the usual
case with YM-matter the abelian multiplets have been eliminated
in \cite{T1} by means of some constraints. In this paper we find the
constraints applying to both the cases of usual and dual supergravity
coupled with the YM-matter. It is demonstrated also that these
constraints are unique ones.

Unfortunately the minimal N=4, D=4 supergravity has some problems
which make it difficult to obtain a realistic model. One
of them is the cosmological term which appears in the lagrangian
where the internal $SU(2)\times SU(2)\sim O(4)$ symmetry
is gauged \cite{FS}. From the ten-dimensional point of view
this gauging corresponds to the $y$-dependent
compactification by Scherk-Schwarz \cite{SS}. Nonminimal
terms can cancel the cosmological term in the action and it is
one of the reasons why they could be important. We hope
that the scheme of reduction described here will be the most
convenient one in the case of nonminimal N=1, D=10 supergravity too.
It is a matter for future speculations.

In section 2 we fix the notations; in section 3 the constraints
are derived; in section 4 the effective potential of the N=4,D=4
supergravity is written.

\section{Notations}

The following index notations are used here:
$$
\begin{array}{ccc}
  \mbox{dimension} & \mbox{flat}         & \mbox{world}           \\
  D=10      & A,B,C,\ldots               & M,N,P,\ldots           \\
  D=4       & \alpha,\beta,\gamma,\ldots & \mu,\nu,\lambda,\ldots \\
  D=6       & a,b,c,\ldots               & m,n,p,\ldots
\end{array}
$$

A number of formulae are taken from the superspace approach where
the 16--component representation for the Majorana-Weil spinors
is the most convenient one. We are not interesting in decomposition
them into four 4--component spinors under the reduction
$D=10 \rightarrow D=4$ here (this procedure has been described
in many papers). So we use the $16\times 16$ $\Gamma$--matrices
with upper and lower indices. The spinorial indices will be omitted
usually.

The fields of pure N=1,D=10 supergravity are given in the introduction;
the fields of the YM-multiplet are: $A_M$ -- the gauge potential
and $\lambda$ -- the gaugino field. They take values in the Lie
algebra of the gauge group G:
$$ A_M=i\,A_M^i\,t^i\;,\ \ \ \lambda=i\,\lambda^i\,t^i\;,     $$
where $t^i$ -- the hermitian G-group generators.

The superspace description of the N=1, D=10 supergravity
is the most convenient one, especially in the nonminimal case.
The superspace (10 ordinary coordinates + 16 spinoral coordinates) has a
nonzero torsion $T_{{\hat A}{\hat B}}{}^{\hat C}$, where ${\hat A}$,
${\hat B}$, ${\hat C}$ take vector or spinoral values.
Due to some set of constraints, defining the field parametrization,
and Bianchi identities the components of the superspace torsion
and curvature are expressed through the fields of the supergravity
multiplet. The field parametrization used here has been introduced in
\cite{N} and slightly modified in \cite{T2,STZ}. This is not the
parametrization with canonical kinetic terms in the lagrangian
\cite{CM,GN} but it is sufficiently convenient one from the superspace
point of view. Nevertheless, the connection with all other
parametrizations can be restored unambiguously if the supersymmetry
transformations are given. In our case they have the form \cite{STZ}:

$$ \delta {E_M}^A = \psi_M\Gamma^A\varepsilon                   $$
$$ \delta \psi_M = \varepsilon_{;\,M} - {1\over 144}\,
(\,3{\hat T}\Gamma_M + \Gamma_M{\hat T}\,)\,\varepsilon -
{1\over 4}\,\Gamma^{PQ}\varepsilon\,S_{MPQ}                     $$
$$ \delta \phi = - \chi\,\varepsilon                            $$
$$ \delta \chi = - {1\over 2}\,\phi_{;\,A}\Gamma^A\varepsilon +
{1\over 36}\,\phi{\hat T}\,\varepsilon +
{1\over 2}\,\Gamma^A\varepsilon\,(\psi_A\,\chi) +
{1\over 2}\,\mbox{Sp}[\Gamma_A\lambda(\lambda\Gamma^A\varepsilon)] $$
$$ \delta M_{M_1 \ldots M_6} = - 3\,\psi_{[M_1}
\Gamma_{M_2 \ldots M_6]}\,\varepsilon
 \ \ \ \ \ \ \ \ \ \ \mbox{(dual)}   $$
$$ \delta B_{MN} = \phi\,\psi_{[M}\Gamma_{N]}\varepsilon -
\,{1\over 2}\,\chi\Gamma_{MN}\varepsilon +
\,{1\over \sqrt{2}}\,\mbox{Sp}(A_{[M}\lambda\Gamma_{N]}\varepsilon)
\ \ \ \ \ \mbox{(usual)}                                        $$
$$ \delta\lambda = -\,{1\over 2 \sqrt{2}}\,{\hat {\cal F}}
\varepsilon                                                     $$
\be
\label{susytr}
\delta A_M = \,{1\over \sqrt{2}}\,\lambda\Gamma_M\varepsilon\;,
\ee
where the semicolon denotes the ordinary covariant
derivative (without torsion);
\be
\label{sabc}
S_{ABC}={1\over 2}\,(2\,\psi_A\Gamma_{[B}\psi_{C]}
+\,\psi_B\Gamma_A\psi_C)\;;
\ee
${\hat {\cal F}}={\cal F}_{AB}\Gamma^{AB}$ ,
${\cal F}_{AB}$ is the superspace matter field-strength:
\be
\label{sfab}
{\cal F}_{AB}=F_{AB}+\,\sqrt{2}\,\psi_{[A}\Gamma_{B]}\lambda\;,
\ee
$F_{AB}$ is the ordinary matter field-strength:
\be
\label{fab}
F_{AB}=\,2\,A_{[B;\,A]}-\,2\,gA_{[A}A_{B]}\;,
\ee
$g$ is the charge; ${\hat T}=T_{ABC}\Gamma^{ABC}$, where $T_{ABC}$
is the superspace torsion with three vector indeces.
It is expressed in a different
way in the cases of usual and dual supergravity:
\be
\label{th}
\phi\,T_{ABC}=-\,2\,H_{ABC}+\,3\,\phi\,\psi_{[A}\Gamma_B\psi_{C]}-
\,3\,\psi_{[A}\Gamma_{BC]}\chi\,+
\,{1\over 2}\,\mbox{Sp}(\lambda\Gamma_{ABC}\lambda)
\ee
in usual case and
\be
\label{tm}
T_{ABC}=\,2\,M_{ABC} -\,{1\over 2}\,\psi^D\Gamma_{DABCE}\psi^E
\ee
in dual. Here
\be
\label{h}
H_{ABC}=\,3\,B_{[AB;\,C]}-\,6\,\mbox{Sp}
(A_{[A}A_{B;\,C]}+\,{2\over 3}\,g\,A_{[A}A_BA_{C]}\,)
\ee
is the usual supergravity field-strength and
\be
\label{m}
M_{ABC}=\,{1\over 6!}\,{\varepsilon_{ABC}}^{D_1\ldots D_7}
M_{D_1\ldots D_6;\,D_7}
\ee
the field-strength of the dual supergravity. The transition from
flat indices to world ones is fulfilled by means of the 10-bein
${E_M}^A$.

By means of the O(1.9) rotation over the flat index one
can vanish, as usual, the ${E_m}^\alpha$ --component of
the 10-bein:
\be
\label{bein}
{E_M}^A=\left(\begin{array}{cc}
             {E_\mu}^\alpha  & {E_\mu}^a \\
             {E_m}^\alpha    & {E_m}^a
             \end{array} \right)=
        \left(\begin{array}{cc}
             {e_\mu}^\alpha  & {\cal B}_\mu^n {e_n}^a \\
             0               & {e_m}^a
             \end{array} \right)
\ee

The Scherk-Schwarz compactification procedure \cite{SS} used here.
It means that any tensor with 4-indices and flat 6-indices is
independent of the coordinates $y^m$ of the internal manifold
but the tensors with world 6-indices depend on $y^m$ in
the following way:
\be
\label{ydep}
{V_{m\ldots}}^{n\ldots}\,(x,y)={{V^{(0)}}_{p\ldots}}^{q\ldots}\,(x)
\,{U^p}_m\,(y)\ldots{{\ov U}^{\,n}}_q\,(y)\ldots \;,
\ee
where ${{\ov U}^{\,m}}_n$ is inverse to ${U^m}_n$. (We shall see,
however, that some exceptions to this rule are needed.) Due to
~(\ref{ydep}) the $y$ -dependence appears in the physical
formulae only in the form:
\be
\label{const}
{C^k}_{lm}=\,2\,{{\ov U}^{\,p}}_l\,{{\ov U}^{\,q}}_m\,
\partial_{[q}{U^k}_{p]}
\ee
Consequently it is necessary to require
that all ${C^k}_{lm}$ must be constants. Then they are the structural
constants of some group with generators
$L_m={{\ov U}^{\,n}}_m\partial_n$
\be
\label{ygroup}
[L_m,\,L_n]={C^p}_{mn}L_p
\ee
and hence obey the Jacoby identity:
\be
\label{jacoby}
{C^p}_{q[k}\,{C^q}_{lm]}=\,0
\ee
In different expressions they enter usually in the
following combination with $U$-matrices:
\be
{{\ov C}^k}_{lm}\equiv{{\overline U}^{\,k}}_n\,
{C^n}_{pq}\,{U^p}_l\,{U^q}_m
\ee

\section{Constraints}

We start from the search of a constraint in the fermionic sector
because it is simpler than bosonic one. Moreover, the fermionic
constraint has the same form both in usual and dual
supergravity while  bosonic constraints have not.
As we have seen in introduction there are
24 Majorana spinors $\psi_m$ in $M^4$ which don't
enter in the multiplet of N=4, D=4 supergravity. So they
must be eliminated by means of a condition like that:
\be
\label{psim}
\psi_m+\,a\,\Gamma_m\chi+\,b\,\mbox{Sp}(A_m\lambda)=\,0
\ee

If we don't want to break the supersymmetry algebra than we must
to require the vanishing of the supersymmetry variation of the relation
~(\ref{psim}). One can expand this variation in powers of
$\Gamma$-matrices:
$$ [\,X+X_{(2)}\Gamma^{(2)}+X_{(4)}\Gamma^{(4)}]\,\varepsilon    $$
The vanishing of this expression for arbitrary $\varepsilon$
implies
$$   X=X_{(2)}=X_{(4)}=\,0\;.                                $$
In general case these conditions have only trivial solution.
But there are unique values of $a$ and $b$ such that $X$ and $X_{(4)}$
are equal to zero identically and the only restriction $X_{(2)}=0$
has a nontrivial solution. So let us require that $a$ and $b$
take exactly these values.
Then all the factors in~(\ref{psim}) are fixed unambiguously.
It explains also why terms of any other type are not written in
~(\ref{psim}).

The explicit form of $a$ and $b$ depends on the field
parametrization. In our notations~(\ref{susytr}) the constraint~(\ref{psim})
takes the form:
\be
\label{start}
2\,\phi\,\psi_m-\Gamma_m\chi+\,2\sqrt{2}\,\mbox{Sp}(A_m\lambda)=\,0
\ee
The supersymmetry variation of~(\ref{start}) leads to:
$$
\phi\,{\tilde \omega}_{mAB}+\,{1\over 2}\,\psi_m\Gamma_{AB}\chi-
E_{m[A}(\phi_{;\,B]}-\,\psi_{B]}\chi)+
$$
\be
\label{con}
+\,2\,\mbox{Sp}
(A_m{\cal F}_{AB}-\,{1\over 8}\,\lambda\Gamma_{mAB}\lambda)=\,0\;,
\ee
where ${\tilde \omega}_{mAB}={e_m}^c{\tilde \omega}_{cAB}$
is the superspace spin-connection:
\be
\label{scon}
{\tilde \omega}_{ABC}=\omega_{ABC}+\,{1\over 2}\,T_{ABC}+S_{ABC}\;,
\ee
$\omega_{ABC}$ is the ordinary spin-connection depending only
on the 10-bein, $S_{ABC}$ is given in~(\ref{sabc}).

One can show that the supersymmetry variation of~(\ref{con})
does not lead to any other restrictions at the mass-shell level.
The following formula helps to do it:
$$ \delta{\tilde \omega}_{mAB}=\varepsilon^{\ov \alpha}
 R_{{\ov \alpha}\,mAB}\;,  $$
where $\ov \alpha$ -- spinoral index, $R$ is the supercurvature
from \cite{T2}.

Consequently~(\ref{start}) and~(\ref{con}) are
all the constraints we must impose.
In fact the relation~(\ref{con}) contains the
five independent conditions:
\be
\label{c1}
\partial_\mu[\,\phi\,g_{mn}+\,2\,\mbox{Sp}(A_m A_n)]=\,0
\ee
\be
\label{c2}
[\,\phi\,g_{q(m}+\,2\,\mbox{Sp}(A_q A_{(m})]\,{{\ov C}^q}_{n)p}=\,0
\ee
\be
\label{c3}
K_{m\mu\nu}=-\,2\,\partial_{[\mu}[\,\phi\,{\cal B}^n_{\nu]}g_{mn}+
\,2\,\mbox{Sp}(A_{\nu]} A_m)\,]
\ee
\be
\label{c4}
K_{mn\mu}=-\,[\,\phi\,g_{pq}{\cal B}^q_{\mu}+
\,2\,\mbox{Sp}(A_p A_\mu)\,]\,{{\ov C}^p}_{mn}
\ee
\be
\label{c5}
K_{mnp}=-\,[\,\phi\,g_{qm}+\,2\,\mbox{Sp}(A_q A_m)\,]\,{{\ov C}^q}_{np}
\ee
Where $K_{MNP}={E_M}^A{E_N}^B{E_P}^C K_{ABC}$ denotes the following tensor:
$$
K_{ABC}=\phi\,T_{ABC}-\,3\,\phi\,\psi_{[A}\Gamma_B\psi_{C]}+
\,3\,\psi_{[A}\Gamma_{BC]}\chi\,-
$$
\be
\label{k}
-\,{1\over 2}\,\mbox{Sp}(\lambda\Gamma_{MNP}\lambda)-\,12\,\mbox{Sp}
(A_{[M}A_{N;\,P]}+\,{2\over 3}\,g\,A_{[M}A_NA_{P]}\,)
\ee
Until we don't substitute the explicit expression for $T_{ABC}$
in~(\ref{k}), the constraints~(\ref{c1}) --~(\ref{k}) have the same
form both in the usual and dual supergravity.

The condition~(\ref{c1}) connects the 6-metric with other fields:
\be
\label{6m}
\phi\,g_{mn}=\eta_{mn}-\,2\,\mbox{Sp}(A_m A_n)\;,
\ee
where $\eta_{mn}=\eta_{pq}^{(0)}{U^p}_m {U^q}_n$,
$\eta_{mn}^{(0)}$ is invariant tensor of the group~(\ref{ygroup}).
Consequently it must be the Killing tensor:
\be
\label{kill}
\eta_{mn}^{(0)}=\,-\,{C^p}_{mq}{C^q}_{np}
\ee

So the condition~(\ref{c2}) is fulfilled automatically because
the structural constants are completely antisymmetric over
all indices due to~(\ref{kill}) and~(\ref{jacoby}):
\be
\label{sym}
{\ov C}_{mnp}\equiv\eta_{mq}{{\ov C}^q}_{np}={\ov C}_{[mnp]}
\ee

But the conditions~(\ref{c3}) --~(\ref{c5}) have a different
meaning for the reduction of usual and dual N=1, D=10 supergravity.

In usual supergravity one must use the expressions~(\ref{th}),(\ref{h})
for the torsion $T_{ABC}$. Here the tensor $K_{MNP}$ gets a simple meaning
\be
\label{kb}
K_{MNP}=-\,6\,B_{[MN;\,P]}
\ee
and the equation~(\ref{c3}) may be integrated:
\be
\label{bb}
2\,B_{m\mu}=-\,\phi\,g_{mn}{\cal B}^n_\mu\,-\,2\,\mbox{Sp}(A_m A_\mu)
\ee
Equations~(\ref{c4}),~(\ref{c5}) are transformed to:
\be
\label{db}
\partial_\mu\,B_{mn}=\,0
\ee
\be
\label{bc}
6\,\partial_{[m}B_{np]}={\ov C}_{mnp}
\ee
If the $B_{mn}$ -- component of the potential obeys the condition
~(\ref{bc}) it must depend on the $y$-coordinates in the way
different from~(\ref{ydep}).

The results~(\ref{start}),~(\ref{bb}) --~(\ref{bc})
have the same form as in
\cite{T1} (the different field parametrization used there)
but we take into account all the terms in the formulae, not
only the lowest order in fermionic fields.

The conditions~(\ref{start}),~(\ref{bb}) --~(\ref{bc})
don't contain derivatives (~(\ref{db}) and~(\ref{bc}) may
be easily integrated) and consequently can be imposed at
the lagrangian level. So in order to obtain the N=4, D=4
lagrangian one can substitute them into N=1, D=10 lagrangian
and then express the $B_{\mu\nu}$ through the pseudoscalar
field $B$ by means of the dual transformation. At the level of
bosonic fields it has been done in \cite{T1}.

In dual supergravity $T_{ABC}$ is given in~(\ref{tm}),(\ref{m}) and
the tensor $K_{ABC}$ takes the form:
$$
K_{ABC}=2\,\phi\,M_{ABC}-\,{1\over 2}\,\phi\,
\psi_D\Gamma^{[D}\Gamma_{ABC}\Gamma^{E]}\psi_E\,-
\,3\,\psi_{[A}\Gamma_{BC]}\chi\,-
$$
\be
\label{kdual}
-\,{1\over 2}\,\mbox{Sp}(\lambda\Gamma_{ABC}\lambda)-\,12\,\mbox{Sp}
(A_{[A}A_{B;\,C]}+\,{2\over 3}\,g\,A_{[A}A_BA_{C]}\,)
\ee
Relations~(\ref{c3}) --~(\ref{c5}) define all of the components of the
field-strength $M_{MNP}$ at the the reduction with except of
$M_{\mu\nu\lambda}$: the potential
$M_{m_1\ldots m_6}$, which enter in $M_{\mu\nu\lambda}$,
becomes directly the pseudoscalar field $B$
of the N=4,D=4 supergravity:
$$
M_{m_1\ldots m_6}=\epsilon_{m_1\ldots m_6}B\ \ ,
$$
where $\epsilon_{m_1\ldots m_6}=1$  if
$\{m_1\ldots m_6\}=\{12\ldots 6\}$.

We see, that all the components of the potential $M_{M_1\ldots M_6}$
(with except of $M_{m_1\ldots m_6}$) are expressed through
other fields in a nonlocal manner. It is not a problem at the level
of equations of motion because they contain ohly the field-strenght.
But constraints~(\ref{c3}) --~(\ref{c5}), containing derivatives,
cannot be imposed at the lagrangian level: if we
try to obtain the N=4, D=4 lagrangian substituting them into
N=1, D=10 one we would get a wrong result.

\section{Potential}

Finally we write the potential of the N=4, D=4 theory. It has
been obtained in \cite{T1} for other field parametrization and
therefore we omit many intermediate formulae.

The pseudoscalar field $B$ doesn't form a part of the potential
because the theory is invariant relative to the transformation
$$ B\rightarrow B+C\;,                                   $$
where $C$ is a constant. Consequently in order to obtain the
potential it is necessary to keep the terms only with
$\phi$ and $A_m$ -- fields in the usual
supergravity lagrangian\footnote{As mentioned in previous
section the dual supergravity
is not convenient for this purpose besause constraints
~(\ref{c3}) --~(\ref{c5}) contain derivatives in this case.}:
\be
\label{lag}
L^{N=1,D=10}=\,{1\over 4}\,\phi\,R+\,{1\over 12}\,\phi^{-1}\,H^2
+\,{1\over 4}\,\mbox{Sp}(F^2)
\ee
(it is the famous lagrangian \cite{CM} rewritten in fields used here).

To derive the lagrangian with correctly normalized kinetic terms
we replace our fields by primed ones
$$ {{e_\mu}^\alpha}' = {(\phi E)}^{1/2}\,{e_\mu}^\alpha $$
$$ e^{-2\phi'} = E                                      $$
$$ {A_m}' = A_m                                         $$
\be
\int e'\,L' = \int e\,L \;\;,
\ee
where $e=\det {e_\mu}^\alpha, E=\det {e_m}^a, L$ is the lagrangian,
$$ {g_{mn}}'= \phi\,g_{mn}=\eta_{mn}-\,2\,\mbox{Sp}(A_m A_n)\;\;, $$
and omit all the primes later on.

The scalar field lagrangian has the form:
\be
L=L_T-U
\ee
$L_T$ is the kinetic part:
$$
L_T=\,{1\over 4}\,R+\,{1\over 2}\,\phi_{;\,\mu}\phi^{;\,\mu}+
\,{1\over 2}\,g^{mn}\mbox{Sp}(A_{m;\,\mu}{A_n}^{;\,\mu})+
$$
\be
+\,g^{mp} g^{nq}\mbox{Sp}(A_m A_{n;\,\mu})\mbox{Sp}(A_p {A_q}^{;\,\mu})
\ee
$U$ -- the potential:
$$
U=\,{1\over 16}\,e^{2\phi} \Bigl\{ \,{C^{\,m}}_{np}g^{nq}\left(2\,
{C^p}_{mq}-g_{mr}{C^r}_{qs}g^{sp} \right) - \Bigr. $$
$$
-\,{1\over 3}\,{ \left[ \,C_{mnp}-\,2\,\mbox{Sp}(3\,A_q A_{[m}{C^q}_{np]}
+\,4\,g\,A_{[m}A_n A_{p]})\, \right] }^2 -   $$
\be
\label{u}
\left. -\,4\,\mbox{Sp} \left[ {(A_p{C^p}_{mn}+
\,2\,g\,A_{[m}A_{n]})}^2 \right] \, \right\}
\ee
where
$$ g_{mn}=\eta_{mn}-\,2\,\mbox{Sp}(A_m A_n)\;\;,\ \ \
C_{mnp}=\eta_{mq}{C^q}_{np}\;,                                   $$
$g^{mn}$ is inverse to $g_{mn}$; the contraction of the indices
$m,n,\ldots$ is fulfilled by means of the $g^{mn}$ -- tensor.

But as it was mentioned in \cite{T1}, the potential~(\ref{u}) cannot
lead to a realistic model.

Indeed, in the case $C_{mnp}=0$ this potential is unbounded
from below because $-g^{mn}$ is not positively definite and singular.

In the case $C_{mnp}\neq 0$ there is a field configuration
$(A_m=0)$ where $U$ takes a negative value:
$$ U=\,{1\over 24}\,e^{2\phi}{(C_{mnp})}^2                      $$
It falls down infinitely together with the rise of the vacuum
expectation value $<\phi>$. Moreover, if $<\phi>=0$ by some reasons
and the potential~(\ref{u}) has a minimum, it must lie below zero.
Hence, after spontaneously symmetry breaking the theory gets an
enormous cosmological constant $\sim {M_{Pl}}^2$.

It is obviously that the abelian matter fields, eliminated here,
cannot solve this problem.

\section{Conclusion}

We have described how to choose the N=4, D=4 supergravity degrees
of freedom from the N=1, D=10 supergravity coupled with the
YM-matter. The main problem of this theory is the non-positively
definite potential. It is possible to solve this taking into
account the Chern-Symons term in the field-strength $H$~(\ref{h}).
In this case supersymmetry transformations have nonminimal corrections
and the starting condition~(\ref{start}) breaks the supersymmetry.
Consequently this condition must be modified by adding appropriate
nonminimal terms. But then the constraint~(\ref{con}) becomes the
equation of third order in $T_{ABC}$ and we don't know whether it is
solvable in a nonperturbative way or not.

\section*{Acknowledgements}

The author thanks M.V.Terentjev for the introduction to the
problem and for the help in the solution of it.

\end{document}